\documentclass[11pt, longbibliography]{article} %
\usepackage{graphicx}
\usepackage{fancyhdr}
\usepackage{makeidx}
\usepackage{amssymb,amsmath}
\usepackage{physics}
\usepackage{upquote}
\usepackage[normalem]{ulem}
\usepackage{xcolor}
\usepackage{cite}
\usepackage{enumerate}
\newcommand{\bd}{\begin{document}}
	\newcommand{\ed}{\end{document}}
\newcommand{\bc}{\begin{center}}
	\newcommand{\ec}{\end{center}}
\newcommand{\vs}{\vspace}
\newcommand{\hs}{\hspace}
\newcommand{\beq}{\begin{equation}}
\newcommand{\eeq}{\end{equation}}
\newcommand{\beqs}{\begin{eqn*}}
	\newcommand{\eeqs}{\end{eqn*}}
\newcommand{\bq}{\begin{quote}}
	\newcommand{\eq}{\end{quote}}
\newcommand{\lb}{\linebreak}
\newcommand{\mb}{\makebox}
\newcommand{\fb}{\framebox}
\newcommand{\mc}{\multicolumn}
\newcommand{\ben}{\begin{enumerate}}
	\newcommand{\een}{\end{enumerate}}
\newcommand{\bit}{\begin{itemize}}
	\newcommand{\eit}{\end{itemize}}
\newcommand{\ov}{\overline}
\newcommand{\un}{\underline}
\newcommand{\lt}{\left}
\newcommand{\rt}{\right}
\newcommand{\ba}{\begin{array}}
	\newcommand{\ea}{\end{array}}
\newcommand{\beqa}{\begin{eqnarray}}
\newcommand{\eeqa}{\end{eqnarray}}
\newcommand{\beqas}{\begin{eqnarray*}}
	\newcommand{\eeqas}{\end{eqnarray*}}
\newcommand{\bfg}{\begin{figure}}
	\newcommand{\efg}{\end{figure}}
\newcommand{\pad}{\partial}
\newcommand{\nn}{\nonumber}
\newcommand{\la}{\leftarrow}
\newcommand{\ra}{\rightarrow}
\newcommand{\lgla}{\longleftarrow}
\newcommand{\lgra}{\longrightarrow}
\newcommand{\La}{\Leftarrow}
\newcommand{\Ra}{\Rightarrow}
\newcommand{\Lra}{\Leftrightarrow}
\newcommand{\Lgla}{\Longleftarrow}
\newcommand{\Lgra}{\Longrightarrow}
\renewcommand{\a}{\alpha}
\renewcommand{\b}{\beta}
\newcommand{\g}{\gamma}
\newcommand{\G}{\Gamma}
\renewcommand{\d}{\delta}
\newcommand{\D}{\Delta}
\newcommand{\e}{\epsilon}
\newcommand{\eps}{\epsilon}
\newcommand{\s}{\sigma}
\renewcommand{\l}{\lamda}
\newcommand{\m}{\mu}
\newcommand{\n}{\nu}
\renewcommand{\S}{\Sigma}
\newcommand{\p}{\pi}
\newcommand{\om}{\omega}
\newcommand{\Om}{\Omega}
\newcommand{\tri}{\triangle}
\newcommand{\ti}{\times}
\newcommand{\f}{\frac}
\newcommand{\ds}{\displaystyle}
\newcommand{\bm}[1]{\mb{{\boldmath $#1$}}}
\newcommand{\alter}[2]{\lt\{ \ba{ll}#1 \\ #2 \ea \rt.}
\newcommand{\alt}[4]{\lt\{ \ba{ll}#1 & \mb{if \, \,}#2 \\ #3 & \mb{}#4 \ea
	\rt.}
\newcommand{\altn}[4]{\lt\{ \ba{rl}#1 & \mb{if \, \,}#2 \\ #3 & \mb{}#4 \ea
	\rt.}
\newcommand{\altif}[4]{\lt\{ \ba{ll}#1 & \mb{if \, \,}#2 \\ #3 &
	\mb{if \, \,}#4 \ea \rt.}
\newcommand{\altnif}[4]{\lt\{ \ba{rl}#1 & \mb{if \, \,}#2 \\ #3 &
	\mb{if \, \,}#4 \ea \rt.}
\newcounter{algc}
\newcounter{romc}
\newcounter{Alphc}
\newcommand{\bl}{\begin{list}{{\it Step} ~\arabic{algc}~:} {\usecounter{algc}
			\setlength{\topsep}{0pt} \setlength{\itemsep}{0pt}}}
	\newcommand{\el}{\end{list}}
\newcommand{\blr}{\begin{list}{~\roman{romc}~:} {\usecounter{romc}
			\setlength{\topsep}{0pt} \setlength{\itemsep}{0pt}}}
	\newcommand{\elr}{\end{list}}
\newcommand{\bla}{\begin{list}{~\Alph{Alphc}~:} {\usecounter{Alphc}
			\setlength{\topsep}{0pt} \setlength{\itemsep}{0pt}}}
	\newcommand{\ela}{\end{list}}
\newcommand{\tsup}{\textsuperscript}
\newcommand{\tsub}{\textsubscript}

\newtheorem{theorem}{Theorem}
\begin{document}

%\preprint{APS/123-QED}

\title{Gate-tunable trion switch for excitonic device applications}
\author{Sarthak Das$^1$, Sangeeth Kallatt$^2$, Nithin Abraham$^1$ and Kausik Majumdar$^{1*}$\\
$^1$Department of Electrical Communication Engineering, \\Indian Institute of Science, Bangalore 560012, India\\
$^2$Center for Quantum Devices, Niels Bohr Institute, University of Copenhagen, Denmark\\
%$^{||}$These authors contributed equally,\\
$^*$Corresponding author, email: kausikm@iisc.ac.in}
%\date{}
\maketitle
{\abstract Trions are excitonic species with a positive or negative charge, and thus, unlike neutral excitons, the flow of trions can generate a net detectable charge current. Trions under favourable doping conditions can be created in a coherent manner using resonant excitation. In this work, we exploit these properties to demonstrate a gate controlled trion switch in a few-layer graphene/monolayer WS\tsub2/monolayer graphene vertical heterojunction. By using a high resolution spectral scan through a temperature controlled variation of the bandgap of the WS\tsub2 sandwich layer, we obtain a gate voltage dependent vertical photocurrent strongly relying on the spectral position of the excitation, and the photocurrent maximizes when the excitation energy is resonant with the trion peak position. Further, the resonant photocurrent thus generated can be effectively controlled by a back gate voltage applied through the incomplete screening of the bottom monolayer graphene, and the photocurrent strongly correlates with the gate dependent trion intensity, while the non-resonant photocurrent exhibits only a weak gate dependence - unambiguously proving a trion driven photocurrent generation under resonance. We estimate a sub-100 fs switching time of the device. The findings are useful towards demonstration of ultra-fast excitonic devices in layered materials.}
\newpage

\section*{Introduction:}
The monolayer semiconducting transition metal dichalcogenides (MoS\tsub2, WS\tsub2, MoSe\tsub2, and WSe\tsub2) exhibit strongly bound two-dimensional excitons with a binding energy on the order of few hundreds of meV, making these ultra-thin monolayers an excellent test bed for excitonic manipulation even at room temperature \cite{ mak2013nmat,plechinger2016trion,jadczak2019room,unuchek2018room,unuchek2019valley}. The neutral excitons ($X^0$) show excellent valley polarization and valley coherence properties that can be readily probed through initialization by circularly and linearly polarized photons, respectively, followed by detection through a circular or linear analyzer \cite{singh2014coherent,gao2016valley,singh2016trion,hao2017trion,wang2014valley,hao2016coherent}. However, controlling these excitonic states electrically remains a challenge due to the charge neutral nature of these excitons. In addition, transport of the exciton also remains challenging due to the ultra-fast radiative recombination of exciton\cite{ceballos2014ultrafast,sun2014observation,hao2016direct,moody2016exciton,robert2016exciton} resulting from the high oscillator strength\cite{palummo2015exciton,gupta2019fundamental,wang2016radiative} - limiting the application of excitonic devices. Recently, this problem has been addressed by creating inter-layer exciton\cite{das2019layer,miller2017long,nagler2017interlayer,yu2015anomalous,okada2018direct} to suppress the fast radiative decay, and exciton transport over several micrometer in the plane of the layered material has been demonstrated \cite{unuchek2018room}. An external gate control has also been achieved by modulating the binding energy of the neutral exciton \cite{unuchek2018room,unuchek2019valley}.

In this regard, the charged exciton or trion ($X^-$) is promising since its intensity can be readily controlled electrically by modulating the doping density using a gate voltage. In addition, the trion, while being optically initiated, can be electrically detected in a spatially nonlocal manner through measuring a charge current \cite{kallatt2019interlayer}. The relatively longer radiative lifetime of trion compared to intra-layer exciton \cite{singh2014coherent,hao2016coherent} further helps in nonlocal detection. Trions can be valley polarized when initiated through a circularly polarized light \cite{gao2016valley}. Trion can also carry valley coherence information when resonantly initialized through linearly polarized light \cite{hao2017trion}. Thus trions are excellent candidates for gate controlled excitonic device applications and also for transferring the valley information into electrical domain. In this work, we demonstrate a fast, gate- and light-tunable vertical trion switch where the trion is coherently initialized through resonant excitation followed by ultra-fast inter-layer transfer and electrical detection - thus generating a gate controlled photocurrent governed by inter-layer trion transport.
\section*{Results and Discussions:}
\begin{figure}[!hbt]
		\centering
		%\vs{-0.1in}
		%\hs{-1in}
		\includegraphics[scale=0.4]{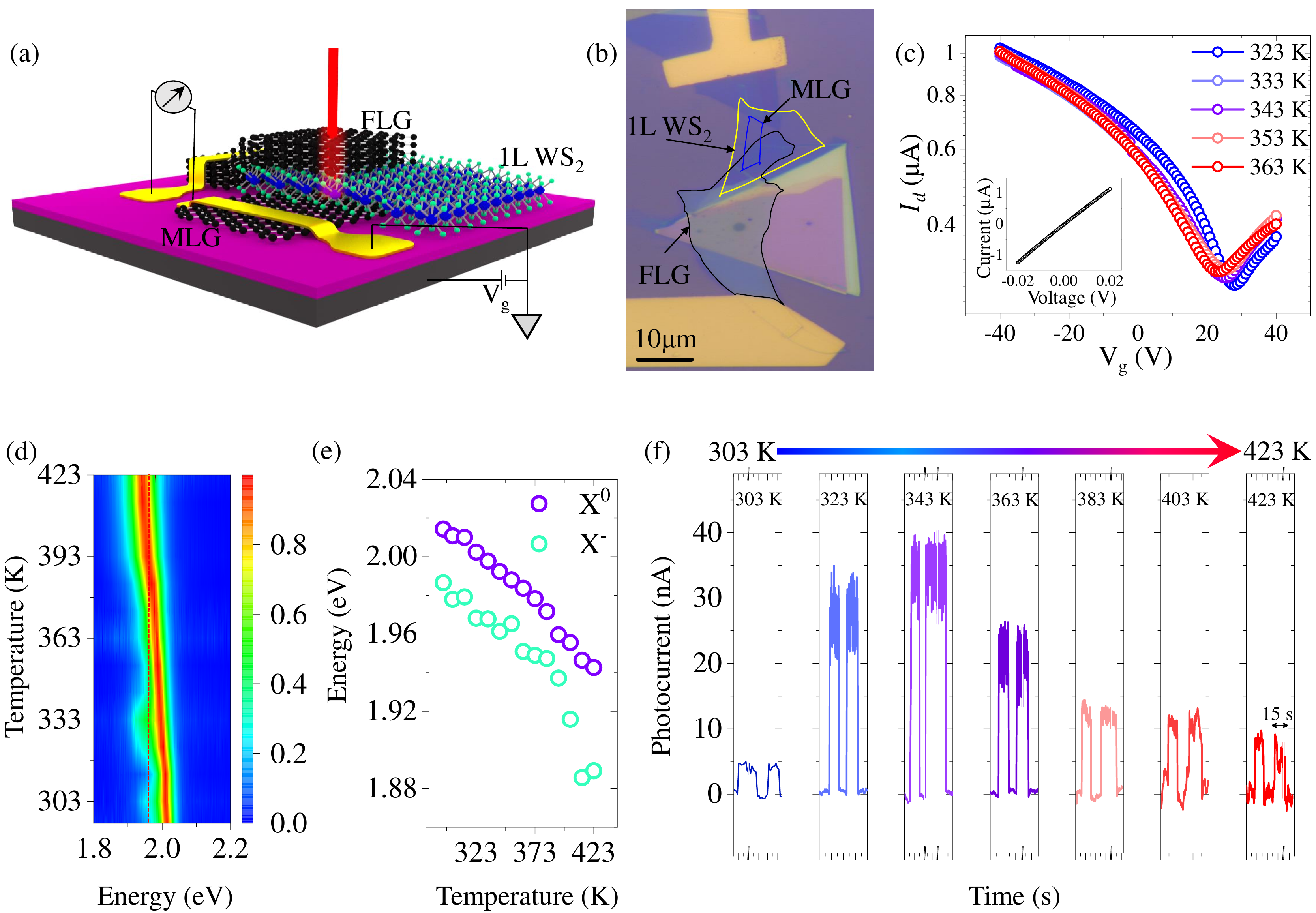}
		%\vspace{-1.8in}
		\caption{\textbf{Photocurrent in FLG/monolayer WS\tsub{2}/MLG vertical heterojunction.}
		(a) Schematic of the experimental device structure where 1L-WS\tsub2 is sandwiched between monolayer graphene (MLG) in the bottom and few-layer graphene (FLG) on the top.
		(b) The optical micrograph of the heterostructure where the MLG, 1L-WS\tsub2 and the FLG are marked with blue, yellow and black outline, respectively.
		(c) The transfer characteristics of the device under dark condition at different temperatures with $V_d=20$ mV. Inset: The dark current-voltage characteristics at 293 K with $V_g=0$ V.
		(d) The color plot of temperature dependent PL spectra with $2.33$ eV laser excitation. The bandgap decreases monotonically with an increase in temperature. The vertical red dashed line corresponds to $1.9591$ eV excitation.
		(e) Variation of exciton ($X^0$) and trion ($X^-$) peak position as a function of temperature showing enhanced separation (and hence higher trion stability) between the two at higher temperatures.
		(f) The transient photocurrent at zero external bias with $1.9591$ eV laser excitation at different temperatures starting from $303$ K to $423$ K in steps of $20$ K.}
\label{fig:schematic}
\end{figure}

Figure \ref{fig:schematic}a shows the schematic diagram of the vertical switch with a monolayer WS\tsub{2} sandwiched between monolayer graphene (MLG) at the bottom and few-layer graphene (FLG) on the top. The heterostructure is prepared on a heavily doped Si substrate coated with $285$ nm thick thermally grown SiO\tsub{2}. The two electrical contacts are deposited to the top and the bottom graphene. More details of the fabrication process are provided in the Methods section. Figure \ref{fig:schematic}b shows the optical image of the device delineating the different stacked layers by marking with different colors.

The fabricated device is then placed on a thermal stage and the terminals of the device are connected to Keithley 2636B SMUs through micromanipulators for electrical measurements. The temperature ($T$) of the device is then increased from room temperature to 423 K in steps of 10 K. At every temperature, the device is illuminated with a linearly polarized laser beam with photon energy of $2.33$ and $1.9591$ eV through an objective with numerical aperture of $0.5$ at different back gate voltages ($V_g$) ranging from $-40$ to $+40$ V. We record the \emph{in situ} photoluminescence (PL) spectra and photocurrent at each $V_g$ and $T$ steps. During measurement, the incident laser power is kept below 8 $\mu$W to avoid any unwanted degradation of the device due to laser induced heating.

Figure \ref{fig:schematic}c shows the transfer characteristics ($I_{dark}$-$V_{g}$) of the vertical device under dark condition at 295 K, when a $V_d = 20$ mV bias is applied across the two terminals. The graphene-like ``V" shaped curve suggests that the $V_g$ dependence primarily arises from the modulation of the chemical potential in the bottom monolayer graphene and the junction acts like a tunneling series resistance. In the inset, the $I_{dark}$-$V_{d}$ plot is shown at $V_g=0$ V confirming the ohmic nature of the junction arising from strong carrier tunneling between the top and the bottom graphene layers through the bandgap of WS\tsub2.

The difference in doping induced workfunction between the top FLG and the bottom MLG results in an asymmetry in the device along the vertical direction, which results in a built-in electric field, allowing a detectable photocurrent ($I_{ph}=I_{light}-I_{dark}$) between the two electrodes under $V_d = 0$ V. This helps us to reduce the average dark current to zero, suppressing the dark noise of the switch dramatically. The sign of the zero-bias vertical $I_{ph}$ at different $V_g$ suggests a net electron flow from the top FLG to the bottom MLG.

Figure \ref{fig:schematic}d depicts the temperature dependent variation of the $A_{1s}$ neutral exciton ($X^0$) peak position as obtained using $2.33$ eV excitation. The individual spectra at each temperature are shown in Supplemental Material S1 \cite{sarthak2020gate}. The $X^0$ and trion ($X^-$) peak positions, as obtained from a Voigt fitting, are plotted as a function of temperature in Figure \ref{fig:schematic}e. One could readily identify that apart from a red shift of the peak position due to temperature induced reduction in bandgap, the trion peak survives up to the highest temperature ($423$ K) used in the experiment - suggesting highly stable trion on the junction. This is further supported by an enhancement in the trion dissociation energy (separation between the $X^0$ and $X^-$ peaks) at higher temperature in Figure \ref{fig:schematic}e, arising from enhanced doping of monolayer WS\tsub2 at higher temperature \cite{kallatt2019interlayer}.

The dashed vertical line in Figure \ref{fig:schematic}d indicates the spectral position of the $1.9591$ eV excitation. This suggests that using a fixed excitation at $1.9591$ eV, with a change in the sample temperature, we can perform a high resolution spectral scan around the $X^0$ and $X^-$ overlap region.
The transient response of the zero-bias $I_{ph}$ under $1.9591$ eV excitation is shown at different temperatures in Figure \ref{fig:schematic}f. Since $V_d=0$, the dark current is ideally zero and practically only limited by the noise of the measurement setup. The vertical jumps in $I_{ph}$ when excitation source is toggled between on and off states indicate that the device works as a fast photonic switch. The magnitude of $I_{ph}$ exhibits a strong non-monotonic trend with temperature, peaking at $T=343$ K, which corresponds to the resonant condition between the excitation energy and the $X^-$ peak. This strongly points to the fact that coherently excited trions participate in the detected photocurrent.

\begin{figure}[!hbt]
		\centering
		%\vs{-0.1in}
		%\hs{-1in}
		\includegraphics[scale=0.45]{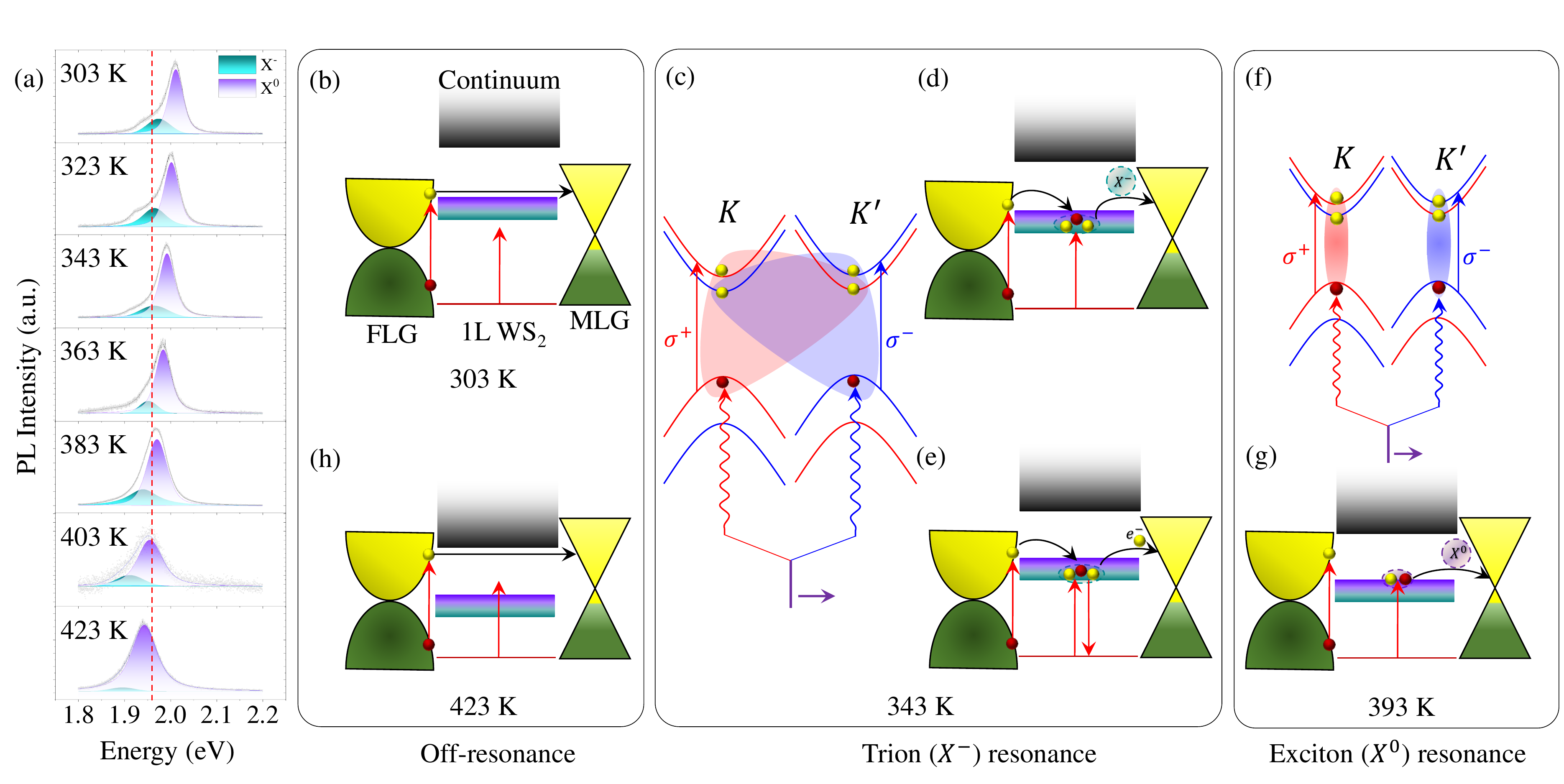}
		%\vspace{-1.8in}
				\caption{\textbf{Resonant trion mediated $I_{ph}$ generation with $1.9591$ eV excitation.}
		(a) Temperature dependent PL spectra of the heterojunction with $2.33$ eV excitation. The $X^-$ and $X^0$ energy states, fitted with two voigt curves (cyan and violet, respectively) are shown separately. The vertical red dashed line shows the spectral position of $1.9591$ eV excitation. The $X^-$ state comes in resonance with $1.9591$ eV excitation at $343$ K while for the $X^0$, it is around $393$ K (not shown in the figure).
		(b-h) Schematic of the photocurrent generation mechanism with $1.9591$ eV excitation at different temperatures. The off-resonance condition is shown in (b) and (h). (b) shows the situation at $303$ K with the excitation is in the sub-optical-bandgap range of 1L WS\tsub2 (at $303$ K, the optical bandgap is $2.014$ eV), while (h) shows the situation at $423$ K where the excitation is above the optical bandgap. The excitation is in resonance with the $X^-$ at $343$ K (c-e).  (c) shows the coherent formation of bright inter-valley trion with linear polarization. The resonantly formed $X^-$ can either be transferred directly to the bottom MLG [shown in (d)] or it can recombine radiatively, releasing an electron, which in turn get transferred to the bottom MLG [shown in (e)]. Both of the processes generate photocurrent, however, the later process is of weaker efficiency. (f-g) show the $X^0$ resonance condition at $393$ K, which does not contribute to the photocurrent generation due to charge neutral nature of $X^0$.}

\label{fig:mechanism}
\end{figure}

The temperature dependent non-monotonic photocurrent generation mechanism is explained in Figure \ref{fig:mechanism}. When the temperature is around $300$ K (top panel of Figure \ref{fig:mechanism}a), the bandgap of WS\tsub2 is higher, and the excitation is well below the $X^0$ and $X^-$ position, thus the resulting photocurrent is weak. The source of this photocurrent from such non-resonant excitation is the photo-excited electrons from the top FLG being driven to the bottom MLG tunneling through the WS\tsub2 barrier by the built-in electric field, as schematically depicted in Figure \ref{fig:mechanism}b. Note that if the temperature changes the relative doping between the top and the bottom graphene, the resulting change in the built-in field would in turn cause a monotonic change in the photocurrent with an increase in temperature. The observed non-monotonicity in $I_{ph}$ magnitude with temperature thus hints a separate mechanism other than just photoelectron tunneling must contribute to the non-monotonicity.

As the sample temperature reaches a value of around $343$ K, the linearly polarized photon creates trions in a coherent manner in $K$ and $K^\prime$ valleys, as explained in Figure \ref{fig:mechanism}c \cite{hao2017trion,hao2016coherent}. Each of the trion consists of a bright exciton in one valley electrostatically bound with an electron in the lower spin-split conduction band from the opposite valley. Since inter-layer transfer is ultra-fast ($\sim$sub-ps \cite{wang2016role,zheng2018phonon,ceballos2014ultrafast,hong2014ultrafast}), which is faster than trion radiative decay ($\sim$tens of ps \cite{wang2014valley,wang2016radiative,hao2017trion,plechinger2016trion,godde2016exciton,mouri2013tunable,lundt2017valley,}), the whole negatively charged trion can be transferred to the bottom monolayer graphene driven by the built-in electric field, as illustrated in Figure \ref{fig:mechanism}d. To account for charge neutrality, the top layer graphene injects an electron to the WS\tsub2, completing the circuit. Thus the trion state acts as an intermediate state to provide a favourable path for the flow of the charge current enhancing the photocurrent. Under resonance, this is the dominating photocurrent transport mechanism in the vertical heterojunction. We also note that after resonant excitation, even if the trion recombines radiatively before being transferred to the bottom graphene layer, the released electron in the conduction band of WS\tsub2 can still be driven to the bottom MLG through the built-in field to generate the photocurrent, as depicted in Figure \ref{fig:mechanism}e. However, noting that the inter-layer transfer process is faster than the radiative lifetime, the latter process of photocurrent is less dominant.

As temperature increases further, the resonance condition breaks and thus the photocurrent also lowers. Around $393$ K , the excitation comes in resonance with $X^0$ (Figure \ref{fig:mechanism}f-g) peak, however the $I_{ph}$ is still quite low. The net charge in the neutral exciton being zero, the  exciton flow does not contribute to the charge current in spite of being resonantly created and transferred to the bottom graphene layer as evidenced from quenching of photoluminescence in several studies \cite{hill2017exciton,froehlicher2018charge,xie2009graphene}. It is also possible that the resonantly created exciton can form trion with emission of phonon, which could eventually generate a charge current. However, the time it takes to form the trion through phonon emission is much longer than inter-layer exciton transfer, suppressing this process, and hence the photocurrent at exciton resonance is also suppressed compared to trion resonance. With further spectral de-tuning through increase in temperature (Figure \ref{fig:mechanism}h), $I_{ph}$ is even more suppressed.

\begin{figure}[!hbt]
		\centering
		%\vs{-0.1in}
		%\hs{-1in}
		\includegraphics[scale=0.5]{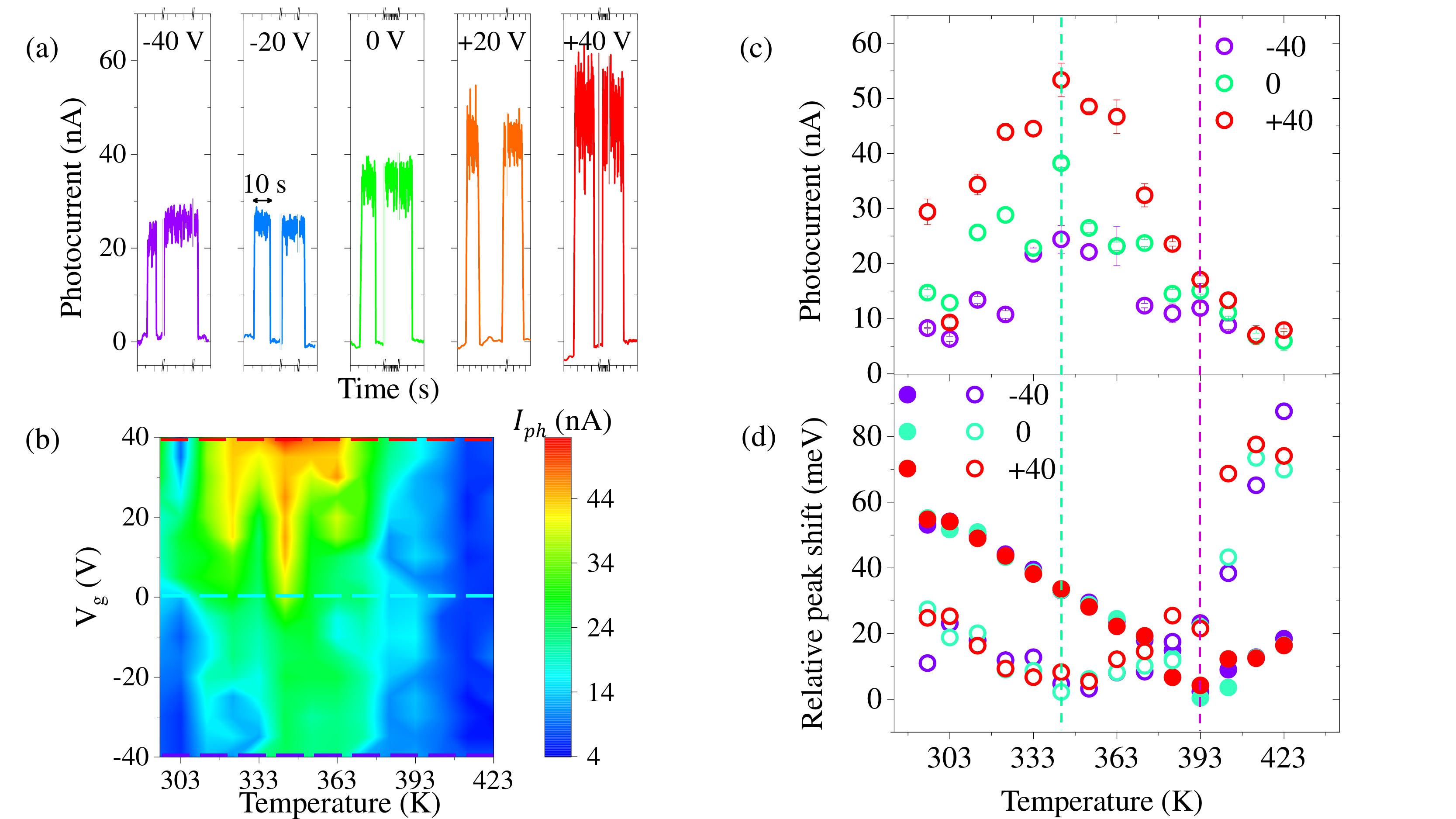}
		%\vspace{-1.8in}
		\caption{\textbf{Gate voltage modulation of generated photocurrent.}
		(a) Modulation of transient photocurrent at $343$ K (trion resonance condition) with different gate voltages ($V_g$) ranging from $-40$ to $+40$ V, showing a monotonic increase in the photocurrent ($I_{ph}$).
		(b) The color plot of $I_{ph}$ for the simultaneous modulation of temperature ($T$) and $V_g$, which reveals stronger $V_g$ modulation of $I_{ph}$ under $X^-$ resonance.
		(c) Horizontal line cuts [shown in dashed lines in (b)] at three different $V_g$ values ($-40$, $0$ and $+40$ V).
		(d) Relative separation of $X^0$ (in solid symbols) and $X^-$ (in open symbols) from $1.9591$ eV excitation at different temperatures The resonance conditions with $X^0$ and $X^-$ are indicated by the vertical dashed lines.}
\label{fig:Iph}
\end{figure}

Figure \ref{fig:Iph}a shows the transient response of $I_{ph}$ with $V_g$ varying from $-40$ to $40$ V, keeping the temperature fixed at $343$ K, suggesting a monotonic increment of $I_{ph}$ with $V_g$. Figure \ref{fig:Iph}b shows the strong tunability of $I_{ph}$ with $T$ as well as $V_g$ through a color plot. With an increase in $V_g$ at the back gate, the WS\tsub2 film can be gated through the bottom MLG film due to its incomplete screening. The gate tunability of $I_{ph}$ is maximum when the excitation is around the resonance with the $X^-$ peak, and tunability reduces on both sides. This is a further evidence regarding the strong role of the trion in the photocurrent generation mechanism. With larger positive $V_g$, formation of $X^-$ is favoured, which in turn increases the photocurrent. The point is further elaborated in Figure \ref{fig:Iph}c-d, by taking horizontal slices from Figure \ref{fig:Iph}b along $V_g = 40, 0,$ and $-40$ V. In Figure \ref{fig:Iph}d, the relative $X^0$ (solid circles) and $X^-$ (open circles) peak shifts with respect to $1.9591$ eV are plotted, with the zero on the vertical axis indicating resonance conditions.

\begin{figure}[!hbt]
		\centering
		%\vs{-0.1in}
		%\hs{-1in}
		\includegraphics[scale=0.5]{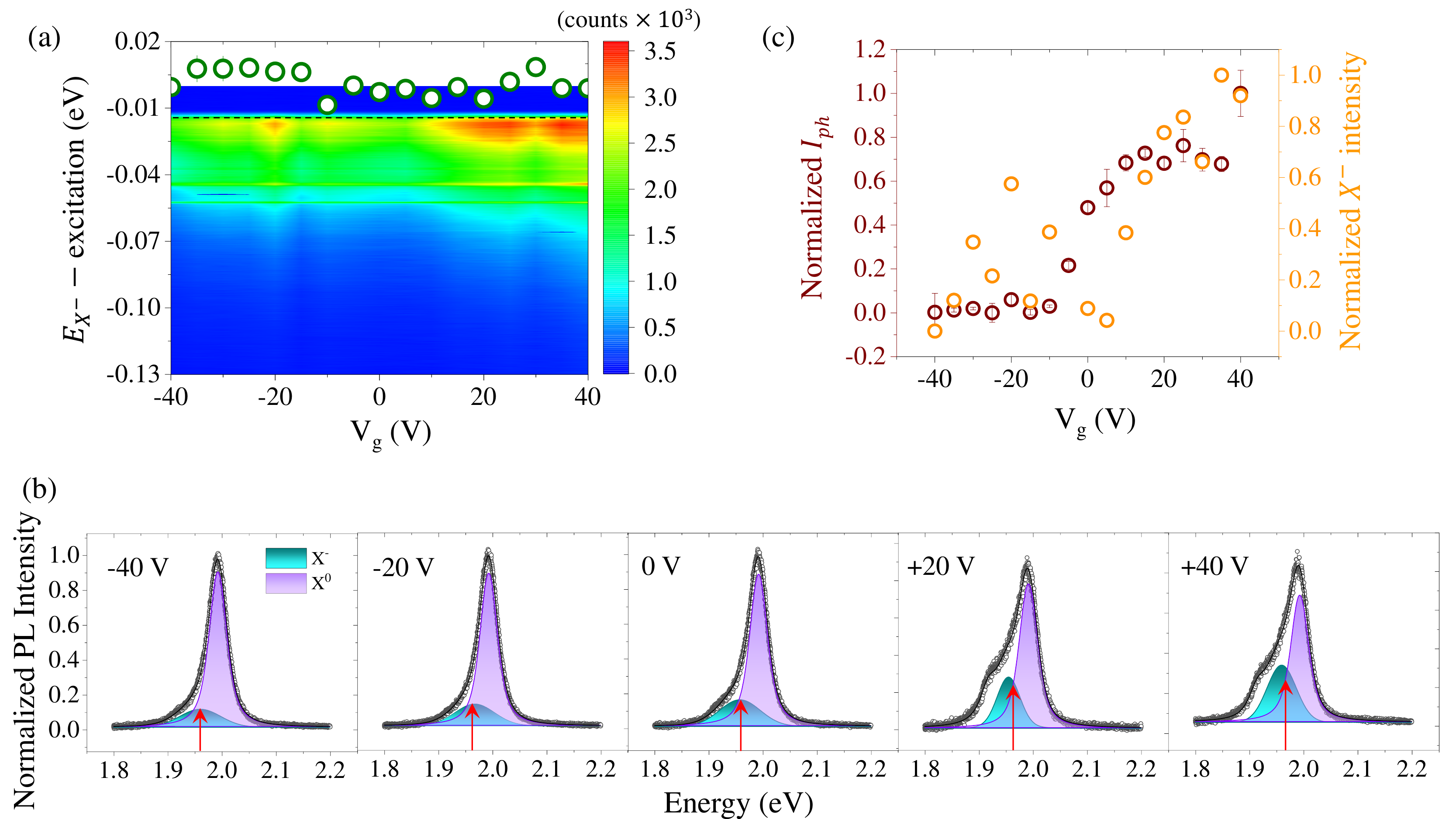}
		%\vspace{-1.8in}
		\caption{\textbf{Correlation of $V_g$ modulation with $X^-$ intensity under resonance.}
		(a) The color plot of the PL intensity obtained with $1.9591$ eV excitation as a function of $V_g$ (in horizontal axis) and the relative emission energy positions of $X^-$ with respect to the excitation energy (in vertical axis). The green open symbols denote the position of the $X^-$ peak obtained from $2.33$ eV excitation.
		(b) Normalized PL spectra with $2.33$ eV excitation at $343$ K showing the $X^-$ and $X^0$ energy states separately at five different $V_g$ conditions. The red arrows indicate the position of the $1.9591$ eV excitation, with which photocurrent is measured.
		(c) Normalized $I_{ph}$ (in brown symbols, left axis) and trion intensity along the horizontal dashed line in (a) (in orange symbols, right axis) for different $V_g$ with $1.9591$ eV excitation at $343$ K, showing strong correlation between the two independent measurements.}
		\label{fig:Vg}
\end{figure}

In order to further justify the point that the observed $I_{ph}$ results from negatively charged  trions, we correlate the photocurrent magnitude with the $X^-$ peak height obtained under $1.9591$ eV excitation. Figure \ref{fig:Vg}a shows in a color plot the photoluminescence intensity around the $X^-$ spectral region as a function of $V_g$. The vertical axis shows the spectral position with respect to the excitation energy. The exact $X^-$ resonance condition is shown by the open circles (as obtained from $2.33$ eV excitation), and cannot be reached during PL measurement with $1.9591$ eV excitation due to the cut-off of the edge filter (individual spectra are shown in Supplemental Material S2)\cite{sarthak2020gate}. Nonetheless, close to the trion resonance, we clearly notice a monotonic increase in the intensity with an increase in $V_g$. The individual spectra with $2.33$ eV excitation at different gate voltages are depicted in Figure \ref{fig:Vg}b, clearly indicating the increasing strength of trion with an increase in the gate voltage. In Figure \ref{fig:Vg}c, we plot the normalized PL intensity as a function of gate voltage along the dashed line ($T=343$ K) in Figure \ref{fig:Vg}a. The normalization of the PL intensity is performed as $\frac{PL(V_g)-PL_{min}}{PL_{max}-PL_{min}}$ where $PL_{min(max)}$ is the minimum (maximum) PL intensity with varying $V_g$ at $343$ K. In the right axis of the same figure, we also plot the normalized $I_{ph}$ [normalized as $\frac{I_{ph}(V_g)-I_{ph,min}}{I_{ph,max}-I_{ph,min}}$ where $I_{ph,min(max)}$ is the minimum (maximum) $I_{ph}$ with varying $V_g$]. The correlation between the two independent measurements is remarkable, unambiguously suggesting the origin of $I_{ph}$ from transport of negatively charged trion in the vertical direction.

Finally, we comment on the estimated switching speed of the trion switch. Since trions are coherently generated through optical excitation, the primary step limiting the intrinsic speed of switching is the inter-layer transfer time of the trion. This inter-layer transfer timescale can be roughly estimated through the difference in homogeneous linewidth broadening of the trion emission between WS\tsub2 at the junction and from WS\tsub2 lying on SiO\tsub2. The total homogeneous linewidth of the trion emission can be estimated from the radiative recombination rate ($\Gamma_r$), non-radiative scattering rate ($\Gamma_{nr}$) and inter-layer transfer rate ($\Gamma_{tr}$) by $\Gamma = \Gamma_r + \Gamma_{nr} + \Gamma_{tr}$. The last term is present only for the heterojunction, and absent for a control WS\tsub2 sample placed on SiO\tsub2. Thus, we have,
\beq
\tau = \frac{\hbar}{2\Gamma_{tr}} = \frac{\hbar}{\Delta \Gamma}
\eeq
where $\Delta \Gamma$ is the difference in the Lorentzian component of the full-width-at-half-maximum of the $X^-$ peaks between the junction and the control sample, after fitting each of them using a Voigt function \cite{gupta2019fundamental,hill2017exciton}. We estimate a value of $\tau \approx 65$ fs suggesting the ultra-fast nature of the trion switch.

In summary, we have demonstrated a gate- and light-controlled trion switch, where trion is optically initiated in a resonant manner and the read out is performed electrically. This can lead to a new paradigm of exciton based optoelectronic switches. The proposed technique exploits the vertical inter-layer transfer of excitonic species, which thus can be extremely fast, compared to relatively slow planar transport of heavy excitons. The efficient controllability through both electrical gating as well as photogating marks an important step towards the realization of trion-based transistor. Since trion can be valley polarized, the valley information can be optically initiated, followed by an electrical sensing, thus transferring the valley information into the electrical domain. The proposed technique can also be used for efficient generation and injection of spin current by resonantly exciting the trion using circularly polarized light - an important step toward realizing spintronic devices.

\section*{Methods}
\textbf{Device fabrication:}
To prepare the heterojunction of MLG/monolayer WS\tsub{2}/FLG we have used dry transfer technique on a highly doped Si substrate covered with $285$ nm thick thermally grown oxide layer. The different layers have been heated subsequently on a hot plate at $70^\circ$ C for 2 minutes in order to get improved adhesion between layers.
	Device contacts are fabricated using standard nanofabrication methods. The substrate is spin coated with PMMA 950C3 and baked on a hot plate at $180^\circ$ C for 2 minutes. This is followed by e-beam lithography with an acceleration voltage of 20 KV, an electron beam current of 220 pA, and an electron beam dose of 200 $\mu$Ccm\tsup{-2}. Patterns are developed using MIBK:IPA solution in the ratio 1:3. Later samples are washed with IPA and dried in N\tsub{2} blow. Electrodes are then made by deposition of 10 nm Ni /50 nm Au films using DC magnetron sputtering at $3\times10^{-3}$ mBar and subsequent lift-off by dipping the substrate in acetone for 15 minutes, followed by washing in IPA and N\tsub{2} drying. The oxide at the back side of the wafer is also etched by dilute HF solution.
	\\
\textbf{Photocurrent measurement:}
Devices are kept on a Linkam thermal stage along with a heater underneath. The laser beam (2.33 eV or 1.9591 eV) is focussed through a 50X objective (NA of 0.5) to the heterostructure with a spot size of approximately 2 $\mu$m. The devices are electrically probed using micro manipulators and Keithley 2636B is used as source meter. Temperature of the stage is increased from $295$ K to $423$ K in steps of $10$ K. A gate bias $V_g$ is applied at the Si substrate ranging from $-40$ to $+40$ Volts in steps of $5$ Volts. At each temperature and at each biasing point, photocurrent measurements are carried out and {\it in situ} photoluminescence spectra are obtained.

%\section*{SUPPLEMENTARY INFORMATION}
%Supporting Information is available on (1) temperature dependent PL spectra with $2.33$ eV laser excitation, (2) Back gate voltage dependent PL spectra with $1.9591$ eV laser excitation at $343$ K.

\section*{ACKNOWLEDGMENTS}
This work was supported in part by a grant under Indian Space Research Organization (ISRO), by the grants under Ramanujan Fellowship, Early Career Award, and Nano Mission from the Department of Science and Technology (DST), and by a grant from MHRD, MeitY and DST Nano Mission through NNetRA.
\section*{Competing Interests}
The Authors declare no Competing Financial or Non-Financial Interests.
%\section*{Data Availability}
%Data available on reasonable request from the authors.
%\section*{Author contribution}

	\bibliographystyle{naturemag}
	\bibliography{references}
	\newpage
\newpage
\newpage
\newpage

\end{document}